\documentclass[aps,pre,twocolumn,groupedaddress,showpacs,amsfonts,10pt,%
tightenlines,floatfix]{revtex4-1}

\usepackage[utf8]{inputenc}
\usepackage{graphicx}
\usepackage{amsmath,amssymb}
\usepackage{upgreek}
\usepackage[usenames,dvipsnames]{xcolor}
\usepackage{braket}
\usepackage{cancel}

\graphicspath{{./Abbildungen/}}

\begin{document}

\title{Epidemics with asymptomatic transmission: Sub-critical phase from recursive contact tracing} 

\author{Lorenz Baumgarten}
\email[]{lbaumgarten@itp.uni-bremen.de}

\author{Stefan Bornholdt}
\email[]{bornholdt@itp.uni-bremen.de}

\affiliation{Institut für Theoretische Physik, Universität Bremen, 28759 Bremen, Germany}

\date{\today}

\begin{abstract}
The challenges presented by the COVID-19 epidemic have created a renewed interest in the development of new methods to combat infectious diseases. 
A prominent property of the SARS-CoV-2 transmission is the significant fraction of asymptomatic transmission. 
This may influence the effectiveness of the standard contact tracing procedure for quarantining potentially infected
individuals. 
However, the effects of asymptomatic transmission on the epidemic threshold of epidemic spreading on networks are largely unknown.  
Here we study the critical percolation transition in a simple epidemic network model in the presence of a recursive contact tracing algorithm for instant quarantining. 
We find that, above a certain fraction of asymptomatic transmission, standard contact tracing loses its ability to suppress spreading below the epidemic threshold. However, we also find that recursive contact tracing opens a possibility to contain epidemics with a large fraction of asymptomatic or presymptomatic transmission. 
In particular, we calculate the required fraction of network nodes participating in the contact tracing for networks with arbitrary degree distributions and for varying recursion depths and discuss the influence of recursion depth and asymptomatic rate on the epidemic percolation phase transition. We test and illustrate our theoretical results using numerical simulations on infection trees and networks. 
We anticipate recursive contact tracing to provide a basis for digital, app-based contact tracing tools that extend the efficiency of contact tracing to diseases with a large fraction of asymptomatic transmission. 
\end{abstract} 

\maketitle 

\section{Introduction}
The methods used to fight the spread of the contemporary COVID-19 epidemic have largely been the same as a hundred years ago during the Spanish flu \cite{franchini2020, wheelock2020}. In particular, contact tracing has been used as a standard procedure that is well understood, both analytically and in network modeling approaches \cite{kretzschmar1996, eichner2003, eames2003, fraser2004, kiss2005}. Some early papers even already considered the concept of recursive contact tracing, i.e.\ not only tracing direct contacts but also contacts of contacts and so on \cite{muller2000, klinkenberg2006}. However, lacking a technology to efficiently implement such a procedure, recursive contact tracing has thus far not been used. \\

However, the arrival of the SARS-CoV-2 epidemic, with its high asymptomatic transmission rate and the possibility of pre-symptomatic infections, presents new challenges that need addressing \cite{yu2020, pribylova2020, khailaie2020}. As such, a renewed interest in recursive contact tracing \cite{okolie2020, bulchandani2020, kojaku2020, lambert2020, barlow2020, endo2020}, as well as in digital contact tracing solutions \cite{faggian2020, hellewell2020, hinch2020, kim2020, xia2020, ferretti2020, mclachlan2020, hernandez2020, prasse2020, ho2020, cencetti2020, barrat2020} that could finally enable recursive contact tracing, has emerged in an effort to surpass the methods of a hundred years ago.\\

In this article, we introduce a simple model that considers an epidemic as a percolation problem, as is common in network epidemiology theory \cite{grassberger1983, cardy1985, newman1999, moore2000, pastor2001, newman2002, warren2002, keeling2005, meyers2007,  pastor2015}, , in combination with a recursive contact tracing algorithm operating on the model. We will study the efficacy of recursive contact tracing and characterize the influence of a disease's asymptomatic transmission rate on the model's critical transition. Our model allows for arbitrary recursion depths, as has only been done in \cite{bulchandani2020}, and our results, to the best of our knowledge, are the first to discuss the relationship of recursion depth and asymptomatic infection rate with regard to the critical transition.\\
 
We find a critical value in the fraction of nodes participating in the contact tracing (corresponding to tracing app usage) which depends on the asymptomatic transmission rate of the disease. Further we find a critical (maximum allowed) asymptomatic transmission rate as a function of the algorithm's recursion depth. 
We show that any disease with arbitrary basic reproduction number and finite asymptomatic rate can be stopped by a sufficiently large recursion depth. 
Finally, we validate our calculations using simulations on infection trees and networks with different degree distributions, as degree distribution can have a significant impact on an epidemic \cite{moore2000, pastor2001_2, newman2002, madar2004, keeling2005, lloyd2005, castellano2010, pastor2015}.
Let us now start by defining the model. 

\section{Theory}
\label{SEC:Theory}
We consider an SIR (susceptible, infected, removed) model with N nodes and an arbitrary degree distribution $p(k)$ in which a proportion $\Phi$ of nodes take part in contact tracing (``use a contact tracing app''). Nodes in the network are infected with a virus with symptomatic rate $\Theta$ and basic reproduction number $R_0$. It is known that in such a network, if we fix $R_0$, the disease has a transmissibility 
\begin{equation}
T=R_0 \frac {\braket k} {\braket {k^2} - \braket k}
\label{EQ:R_0}
\end{equation}
\cite{newman2002}. Carriers of the disease will be able to infect their susceptible neighbors with probability $T$ one time step after being infected themselves and be immune and non-contagious afterwards.\\

If an infectious person is symptomatic and uses the contact tracing app, this will trigger an alarm on the app and warn neighboring nodes of the chance of being infected, sending them into quarantine for their one infectious time step so they effectively skip the infectious state and jump directly to the recovered stage.\\

\begin{figure}[hbt]
	\centering
	\includegraphics[width=1\linewidth]{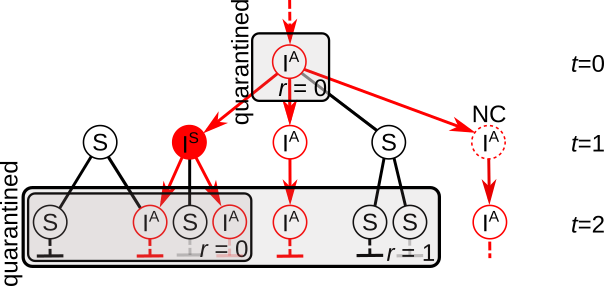}
	\caption{Illustration of the quarantining algorithm with an infection spreading from top to bottom. Nodes with a black outline are not infected (susceptible=S), nodes with a red outline are infected (I), either symptomatically (filled nodes, I$^\text{S}$) or asymptomatically (unfilled nodes, I$^\text{A}$), and nodes with a dashed outline are not using the contact tracing app (non-compliant, NC). Red arrows indicate the spread of the infection, while black lines indicate non-infectious connections between nodes. The time $t$ indicated on the right hand side marks the time at which infected nodes are infectious---or, in the case of the uninfected nodes, the latest time at which the app would consider them to be possibly infectious.
	While nodes could reappear in later time steps, e.g. the node in the $t=0$ row could also be shown in the $t=2$ row as it is connected to (most of) the nodes in the $t=1$ row, we only show nodes once for visual clarity.
	At time $t=1$ a symptomatic node triggers an alarm on the app. For recursion depth $r=0$, only its nearest neighbors are quarantined. These quarantined nodes cannot infect any other nodes, as indicated by the blocked outgoing connections. For $r=1$, the app considers every nearest neighbor of the symptomatic node as a possible origin of the symptomatic node's infection, and therefore quarantines all nodes that the infection could have spread to within two time steps from these nearest neighbors. This results in every node with a distance of exactly three to the symptomatic node being quarantined, so long as the connection is not interrupted by a node not using the app or by a node that was in quarantine itself at its time of infection or in the time step after infection, as shown on the right hand side. This can, of course, also include nodes which have not yet actually been in contact with any infected nodes, as shown by the leftmost nodes in the $t=1$ and $t=2$ rows. Note that, although the infection chain is shown in a tree-like structure for visual clarity, these nodes can be part of a network of arbitrary structure.}
	\label{FIG:Quarantine_Algorithm}
\end{figure}

We can consider higher degrees of recursivity $r$ for the app, meaning how many time steps in the past the app will consider to guess who might currently be infected. For $r=0$, only the node's direct neighbors are sent into quarantine. For $r=1$ in addition to those nodes who are quarantined for $r=0$, any node with a distance of exactly three to the symptomatic node is quarantined, for $r=2$ any node with a distance of five is quarantined, and so on. This is illustrated in Figure \ref{FIG:Quarantine_Algorithm}.
The algorithm disregards any possible immunities due to nodes having already been infected previously, but it does consider breaks in the infection chain that are caused by the app's own quarantining algorithm, i.e., if a node was quarantined at time $t$, the app does not consider this node a possible infection spreader at that time step.\\
Given a vector $\vec S$ of symptomatically infected nodes at time $t_0$,
$$S_i = \begin{cases}
	1 & \text{if node $i$ is symptomatically infected}\\
	0 & \text{otherwise}\end{cases},$$
the vector of nodes $\vec U$ using the app, the vectors $\vec Q(t)$ of quarantined nodes and $P(t)$ of not quarantined nodes at time steps $t\le t_0$, and the adjacency matrix $A$, the vector of quarantined nodes at $t=t_0+1$ can be calculated by
\begin{widetext} 
\begin{align*}
	Q(t_0+1) &= \underbrace{\left\{A\cdot\left[\vec S \cdot \vec U\cdot \vec P(t_0)\right]\right\}\cdot \vec P(t_0) \cdot \vec U}_{r=0}\\
	&+ \underbrace{\left[A\cdot\left(A\cdot\left\{\left[A\cdot\left(\vec S \cdot \vec U\right)\right] \cdot \vec P(t_0-1) \cdot \vec P(t_0-2) \cdot \vec U\right\}\right) \cdot\vec P(t_0-1) \cdot \vec P(t_0) \cdot \vec U\right]\cdot \vec P(t_0) \cdot \vec U }_{r=1}\\
	& + \underbrace{\dots}_{r>1}.\\
\end{align*}
\end{widetext}
Multiplications with $\vec P(\cdot)$ ensure that a considered node in the backtracking chain was neither quarantined at its supposed time of infection nor at the time it could have infected its neighbors, and multiplications with $\vec U$ ensure that all nodes in the backtracking chain use the app.\\
We now calculate the probability that an infected node is correctly put into quarantine by our algorithm. For this, we assume an infinitely large network, with a low enough fraction of the population being infected that an infected node is only quarantined as a result of its own infection chain and not coincidentally swept up in the quarantine caused by a different infection. We assume that the clustering in the network is negligible so that we can consider the infection chain effectively as a tree.\\
For $r=0$, both the infected node and the infecting node must be part of the network and the infecting node needs to be symptomatic.\\
Therefore, a first approximation of the probability $P_q^{r=0}$ of an infected node being correctly put into quarantine is simply
\begin{equation}
P_q^{r=0}(\Phi, \Theta) = \Phi^2\Theta.
\end{equation}
However, because the infecting node cannot have been quarantined, its probability of using the app is
\begin{align*}
	\Phi' &= \frac {\Phi (1-\frac{P_q^{r=0}}{\Phi})}{\Phi(1-\frac{P_q^{r=0}}{\Phi})+(1-\Phi)} \\
	&= \frac{\Phi-P_q^r}{1-P_q^r} \le \Phi,
\end{align*}
as the amount of nodes using the app with the ability to infect other nodes is reduced by a factor $(1-P_q^{r=0})$, and therefore
\begin{equation}
	P_q^{r=0} = \Phi \Phi' \Theta.
\end{equation}
For higher degrees of recursion, the chance of being quarantined is increased
\begin{align}
	P_q^{r>0} &= P_q^{r=0} + \underbrace{(1-P_q^{r=0})\Phi''P_1}_{\text{r}=1} + \underbrace{\cdots}_{r>1} \\
	&= \Phi\Phi'\left[ P_0 + (1-P_0)\Phi''\left\{P_1 + (1-P_1)\Phi''\left(\cdots\right)\right\}\right]\\
	&\text{with }P_0=\Theta
\end{align}
Here, in every part of the sum, the chance of a node having already been quarantined due to a lower recursion level is excluded via $(1-P_i)$, and a factor $\Phi''$ is added for the chance of the next upstream node using the app. The factor $\Phi''$ represents the chance of a node using the app if the next downstream node has not been quarantined, and needs to be used for nodes that are two or more levels above the currently regarded node in the infection tree. The chance of such a node using the app regardless of the behavior of its downstream nodes is $\Phi'$. The chance of a downstream node, which is using the app, of a node that is also using the app not being quarantined is approximately $\left(1-\frac{P_q^r}{\Phi\Phi'}\right)$. Since we assume both infecting node and infected node to be using the app, the factor $\Phi\Phi'$ is removed from $P_q^r$. This approximation disregards that the upstream node not being quarantined also influences the chance of its downstream node being quarantined. Then the chance of an upstream node using the app, given that its downstream node is using the app and has not been quarantined is
\begin{align}
	\Phi'' &= \frac {\Phi'\left(1-\frac{P_q^r}{\Phi\Phi'}\right)}{\Phi'\left(1-\frac{P_q^r}{\Phi\Phi'}\right) + (1-\Phi')}\\
	&= \frac {\Phi\Phi' - P_q^r}{\Phi-P_q^r}.
\end{align}
Next, we need to calculate the chance $P_i$ of a node being quarantined due to the $i$'th recursion step, given that its $r$ nearest upstream nodes are using the app. For simplicity's sake, we start with $P_1$. Here, a leaf node $i$ is quarantined due to the first recursion step if any of the downstream nodes of $i$'s second degree upstream node, which we call $j$, have been infected, use the app, and are symptomatic. The chance of one node fulfilling these conditions is $\Phi'\Theta T$. Since just one node needs to cause an alarm on the app, the chance of being quarantined is
\begin{equation}
	P_1 = 1-(1-\Phi\Theta T)^{n},
\end{equation}
where $n$ is the average number of $j$'s downstream nodes minus one. We subtract one, since one of $j$'s downstream nodes is $i$'s direct upstream node and would already have caused $i$ to be quarantined in the zero'th recursion step, if it were symptomatic. Since the chance of a node of degree $k$ being infected is proportional to $kp(k)$ \cite{madar2004}, the average number of downstream nodes minus one is
\begin{equation}
	n = \frac{\sum_{k=2}^\infty k(k-2)p(k)}{\sum_{k=2}^\infty kp(k)},
\end{equation}
where we subtract two from k because of the one downstream node that is not considered and $j$'s upstream node.
Therefore,
\begin{align}
	P_1 &= 1-(1-\Theta\Phi T)^\frac{\sum_{k=2}^\infty k(k-2)p(k)}{\sum_{k=2}^\infty kp(k)} \\
	= P_1(x)|_{x=2} &= \left.1-(1-\Theta\Phi T)^\frac{\sum_{k=x}^\infty k(k-x)p(k)}{\sum_{k=x}^\infty kp(k)}\right|_{x=2} .
\end{align}
We indicate how many connections are removed when calculating $n$ via the variable $x$. \\
For the second recursion step, at least one of the downstream nodes of $j$'s upstream node, which we call $l$, must fulfill the condition of $P_1$, meaning that any one of their downstream nodes must be infected, using the app, and symptomatic. This chance is given by
\begin{align}
	P_2 &= 1-\left[1-P_1(1)\tilde\Phi\right]^\frac{\sum_{k=2}^\infty k(k-2)p(k)}{\sum_{k=2}^\infty kp(k)} \\
	= P_2(x)|_{x=2} &= \left.1-\left[1-P_1(1)\tilde\Phi\right]^\frac{\sum_{k=x}^\infty k(k-x)p(k)}{\sum_{k=x}^\infty kp(k)}\right|_{x=2} \\
	\text{with } \tilde\Phi &= \frac{\Phi T(1-\Theta)}{\Phi T(1-\Theta)+(1-\Phi T)}.
\end{align}

Here, in $P_1(x)$, we do not discount one of each node's downstream nodes, since these nodes are not upstream nodes of node $i$, and therefore all of their downstream nodes need to be considered. Thus, we use $P_1(1)$ instead of $P_1(2)$. Also, we use $\tilde\Phi$, because nodes that are using the app and symptomatically infected would have already caused a quarantine in a previous time step and can therefore not be part of the considered tree. Similarly, the equation for following recursion steps is
\begin{equation}
	P_i(x) = 1-\left[1-P_{i-1}(1)\tilde\Phi\right]^\frac{\sum_{k=x}^\infty k(k-x)p(k)}{\sum_{k=x}^\infty kp(k)}.
\end{equation}
Summarizing these calculations, the chance of a leaf node being quarantined with a recursion degree of $r$ is
\begin{align}
	P_q^r &\approx \Phi\Phi' \sum_{i=0}^r\left(\left\{\prod_{j=0}^{i-1}\left[1-P_j(2)\right]\Phi''\right\}P_i(2)\right)\label{EQ:P_q}\\
	\text{with } P_i(x) &= \begin{cases} \Theta & \text{if } i=0\\
		1-(1-P_0(1)\Phi T)^{n(x)} & \text{if } i=1\\
		1-(1-P_{i-1}(1)\tilde\Phi)^{n(x)} & \text{otherwise}
	\end{cases}\\
	\text{and } n(x)&=\frac{\sum_{k=x}^\infty k(k-x)p(k)}{\sum_{k=x}^\infty kp(k)}
\end{align}
Note that \eqref{EQ:P_q} is a self-consistent equation, since $\Phi'$ and $\Phi''$ contain $P_q^r$.

\section{Theoretical Results}
\begin{figure}
    \centering
    \includegraphics[width=1\linewidth]{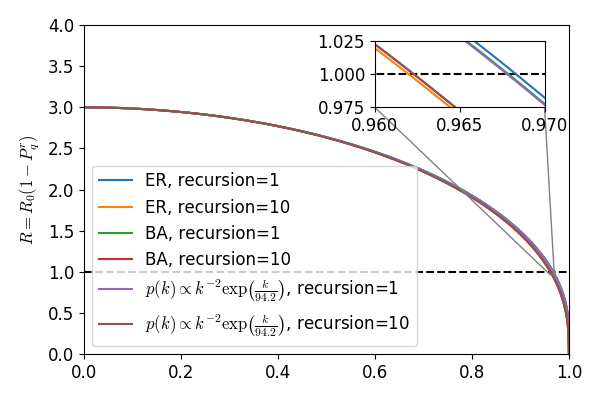}
		\caption{Reduction of the reproduction number $R$ as a function of the app-usage rate $\Phi$ for an Erd\H{o}s-R\'enyi (ER) network, a Barab\'asi-Albert (BA) network with a cutoff $\kappa=1000$, and a scale-free network with exponential cutoff, with $R_0=3$ and $\Theta=0.5$. The dashed line shows the critical value $R_0(1-P_q^r)=1$, and the inset is a blowup around the critical points. The curves of the BA distribution and the scale-free distribution with exponential cutoff lie so close to each other that they cannot even be distinguished in the inset.}
    \label{FIG:Suppression_Lines}
\end{figure}

\begin{figure}
    \centering
    \includegraphics[width=1\linewidth]{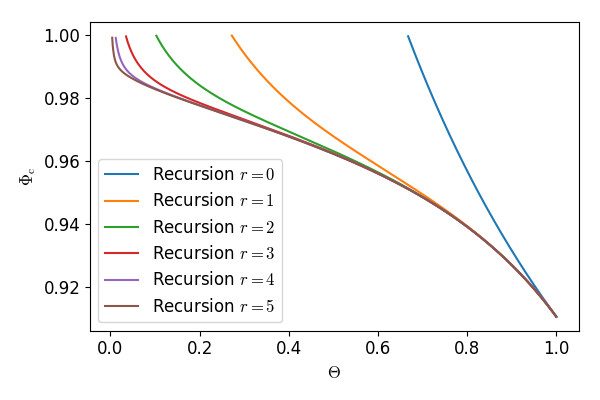}
		\caption{Critical value $\Phi_\text{c}$ as a function of the symptomatic rate $\Theta$ for different recursion depths $r$ with $R_0=3$. Since the ER distribution and the scale-free distribution with an exponential cutoff again yield almost the same results, we only plot $\Phi_\text{c}$ for the Barab\'asi-Albert distribution with average degree $\braket k=4$ and cutoff $\kappa=1000$.}
    \label{FIG:Theta_Crits}
\end{figure}

\begin{figure}
    \centering
    \includegraphics[width=1\linewidth]{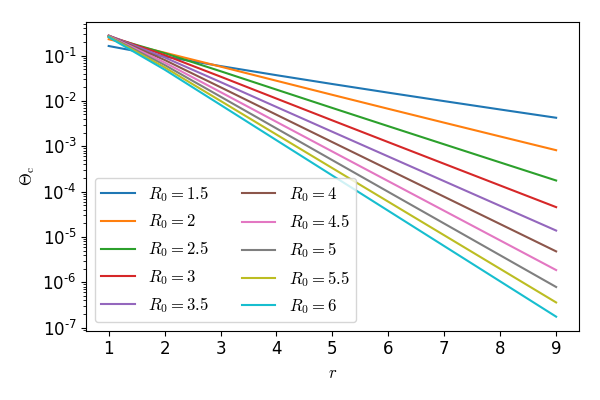}
    \caption{Critical symptomatic rate $\Theta_\text{c}$ below which an epidemic cannot be stopped even for $\Phi=1$ as a function of the recursion depth $r$ for different basic reproduction numbers $R_0$ using a Barab\'asi-Albert distribution with average degree $\braket k=4$ and cutoff $\kappa=1000$. For large recursion depths, the critical value $\Theta_\text{c}\rightarrow 0$ for all basic reproduction numbers, whereas for $r=1$ there is a maximum $\Theta_\text{c}^\text{max}\approx 0.28$ at $R_0\approx 3.6$.}
    \label{FIG:Crit_Thetas}
\end{figure}

It is easy to see that the upper limit of $P_q^r$ is
\begin{equation}
    P_q^r \le \Phi\Phi' < \Phi ~\text{if }\Phi<1,
\end{equation}
so contact tracing by recursive backtracking is strictly worse than vaccinating a fraction $\Phi$ of the population. Since such a vaccination strategy is already insufficient to stop an epidemic on an infinitely large scale-free network with a degree distribution $p(k)\propto k^{-\gamma}$ with $\gamma\le3$ \cite{madar2004}, recursive backtracking can also not stop such an epidemic for $\Phi<1$.\\
However, there is still something that can be learned from taking a closer look at scale-free networks. For $\gamma\le 3$, the sum $\sum_{k=2}^kk^2p(k)$ in the exponent of the $P_i$'s diverges, therefore $P_1\rightarrow 1$ (if $\Phi\Theta T>0$), and $P_q^r$ becomes
\begin{equation}
    P_q^r = \Phi\Phi'\left[\Theta + (1-\Theta)\Phi''\right].
\end{equation}
We can see that all infected nodes that can be caught by the algorithm will already be detected in the first recursion step.\\
Luckily, real world networks are not infinitely large, so the sum mentioned previously will not diverge, so recursive backtracking will be able to stop epidemics for $\Phi<1$. For such networks, we expect the observation made for infinitely large scale-free networks to be still be relevant, i.e., the closer a real world network is to an infinitely large scale-free network, the less will the epidemic threshold $\Phi_\text{c}$ be affected by recursion depths past $r=1$.\\
In Figure \ref{FIG:Suppression_Lines}, we show the reduction of the reproduction number $R=R_0(1-P_q^r)$ as a function of $\Phi$ for different degree distributions and recursion depths. For degree distributions we choose a simple Erd\H{o}s-R\'enyi (ER) network with average degree $\braket k = 4$, a Barab\'asi-Albert (BA) network with average degree $\braket k =4$ and a cutoff at $\kappa=1000$, i.e. $p(k)=0$ for $k>\kappa$, and as a realistic example, a scale-free network with exponential cutoff $p(k)\propto k^{-2}\exp\left(\frac k {94.2}\right)$ that produces an epidemic threshold comparable to that of urban networks for SARS \cite{meyers2005}. We choose the transmissibility $T$ so that all networks have a realistic basic reproduction number for the SARS-CoV-2 virus, $R_0=3$, and we choose $\Theta=0.5$. We see that the degree distribution only has a minuscule effect on $P_q^r$, and that increasing the recursion depth past $r=1$, while having the largest effect for ER networks (outer two lines in the inset), still barely decreases the critical value $\Phi_\text{c}$.\\
We can also calculate the critical value $\Phi_\text{c}$ as a function of the symptomatic rate $\Theta$, as is shown in Figure \ref{FIG:Theta_Crits}.
There is a large visible difference between the classic contract tracing method with $r=0$ and recursive contact tracing, even for relatively large values of $\Theta$. 
While for $r>0$ the recursion depth has little influence on $\Phi_\text{c}$ for large values of the symptomatic rate $\Theta$, we see that there is a critical value $\Theta_\text{c}$, depending on the recursion depth, below which, even with $\Phi=1$, an epidemic cannot be stopped. This critical value is approximately halved when going from the classical method $r=0$ to $r=1$, meaning that recursive contact tracing is an effective method to combat diseases with high asymptomatic rates which would not have been able to be stopped by previous contact tracing methods.\\
The critical value $\Theta_\text{c}$ is shown in Figure \ref{FIG:Crit_Thetas} as a function of the recursion depth for different values of $R_0$. The critical value $\Theta_\text{c}$ exponentially decreases with $r$, with $\Theta_\text{c}\rightarrow 0$ for $r\rightarrow\infty$. Therefore, any disease with a symptomatic rate $\Theta>0$ and arbitrarily large basic reproduction number $R_0$ can be stopped via recursive contact tracing, given a sufficiently large recursion depth and app usage rate.

\section{Simulations}
\begin{figure*}
    \centering
    \includegraphics[width=1\linewidth]{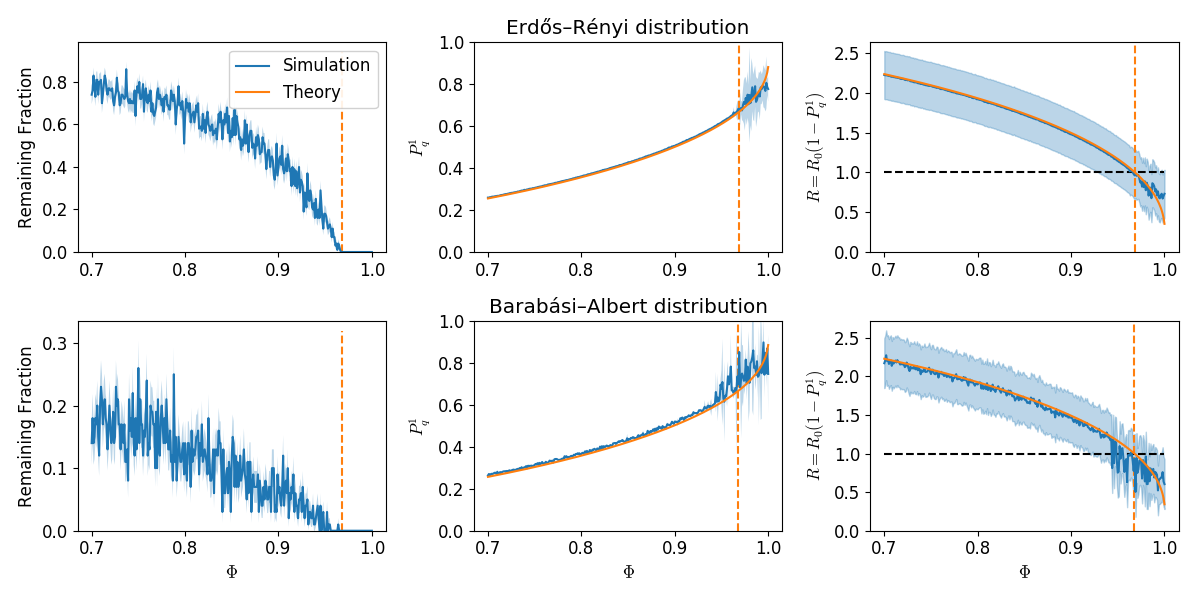}
		\caption{Fraction of trees in which the epidemic survives 100 time steps (left column), probability of an infected node being quarantined $P_q^1$ (center column), and reproduction number $R$ (right column) for trees built with an ER degree distribution (upper row) or a BA degree distribution with cutoff $\kappa=1000$ (lower row), with $r=1$, $R_0=3$, and $\Theta=0.5$. Blue lines show the averages of 100 trees per data point, unbroken orange lines show the theoretical results for $P_q^1$ and $R$, and dashed orange lines show the theoretical critical value $\Phi_\text{c}$. The dashed black lines in the reproduction number diagrams show the critical value of $R$. Note that the measurement for the reproduction number $R$ and the quarantined fraction $P_q^1$ are skewed near or past the critical point, because the measurements here are dominated by just the beginning of the tree where the quarantining algorithm does not have enough history yet to quarantine nodes.}
    \label{FIG:Tree_Simulations}
\end{figure*}

\begin{figure*}
    \centering
		\includegraphics[width=1\linewidth]{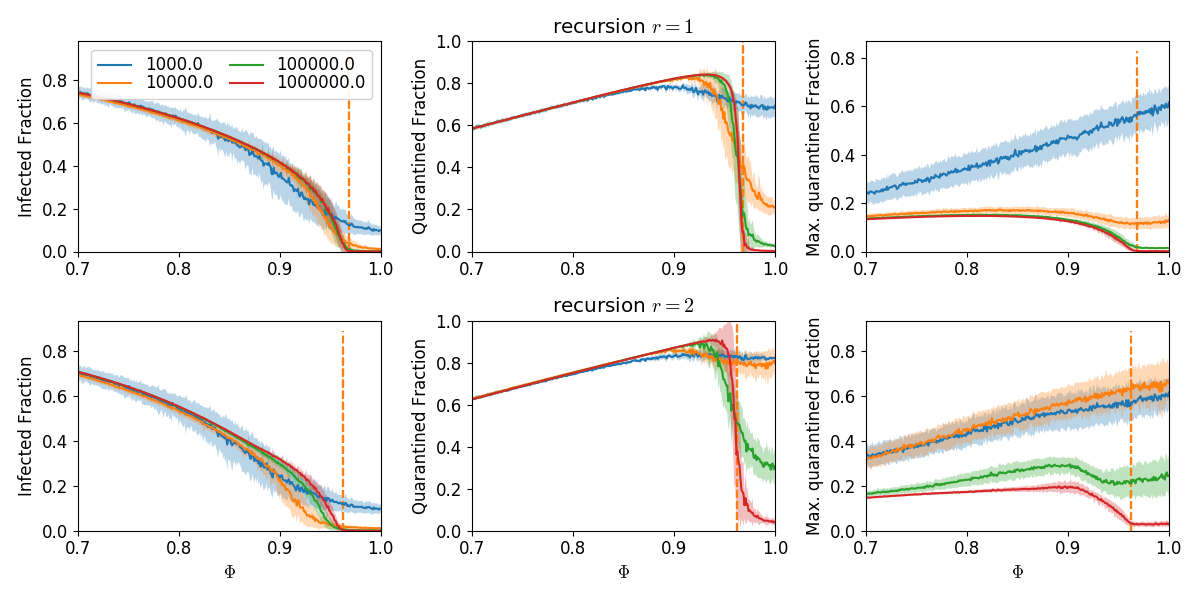}
		\caption{ Fraction of infected nodes (left), fraction of nodes that have ever been quarantined (center), and maximum number of nodes that have been quarantined at one time (right) for ER networks with recursion depth $r=1$ (top row) and $r=2$ (bottom row) as a function of the app-usage rate $\Phi$. Different color graphs show networks of different sizes $N$, and orange dashed lines show the theoretical critical value $\Phi_\text{c}$. All data points are the average of 100 simulation runs. }
		\label{FIG:Network_Simulations}
\end{figure*}

To test the accuracy of our calculations in section \ref{SEC:Theory}, we simulate infection trees with recursive backtracking. The simulation starts with a single infected node, and each time step for each infected, unquarantined leaf node $k-1$ downstream nodes are added, with $k$ proportional to $kp(k)$. These new leaf nodes are infected with probability $T$ and symptomatic with probability $\Theta$. Then, according to the rules described in section \ref{SEC:Theory}, infected leaf nodes may be quarantined, causing them to not receive any downstream nodes. We let these dynamics run for 100 time steps or until there were 10000 new infected leaf nodes added in a time step, at which point we consider the epidemic out of control. In Figure \ref{FIG:Tree_Simulations}, we show the fraction of trees in which the epidemic is not stopped within 100 time steps, the fraction of quarantined nodes, and the average reproduction number $R$ for trees using an ER degree distribution or a BA degree distribution with a cutoff $\kappa=1000$. We see a very good agreement between our calculation and simulations for recursions $r=1$, see Figure \ref{FIG:Tree_Simulations}. We have also verified that our calculations and simulations agree very well for larger recursion depths.
\\
Next, we move away from the tree structure and use networks instead. In these networks, we start with ten initially infected nodes, which are chosen with a probability proportional to $kp(k)$, and we let the dynamics run until no new nodes are infected within a time step. Figure \ref{FIG:Network_Simulations} shows the fraction of infected nodes, the fraction of nodes that have ever been quarantined, and the maximum fraction of nodes that has been quarantined at one point in time for ER networks with different recursion depths.\\
For the network size $N\rightarrow\infty$, we see that the fractions of infected and quarantined nodes drop to zero at the theoretical critical value $\Phi_\text{c}$.
For higher recursion depths and relatively small networks, the infected fraction is already kept quite low below the theoretical critical value because a large fraction of nodes is being quarantined and therefore the assumption we made in section \ref{SEC:Theory} that nodes are not coincidentally swept up in unrelated infection trees does not hold anymore; however, this lower infected fraction comes at the cost of wrongly quarantining a relatively large fraction of nodes. Also, this effect is mitigated for larger network sizes $N$.\\
For BA networks, especially for large networks, the infection dies out quickly even for low values of $\Phi$, because the infection dynamics are dominated by the strongly connected hub nodes which, after some time, will be in the recovered state, and therefore the effective degree distribution for the infection is quickly cut off for larger $k$. Additionally, in a BA network the first few nodes which are added to the network and later are likely to grow into the strongest connected nodes are likely to connect to each other and have common neighbors, meaning that the assumption we made in section \ref{SEC:Theory} of low clustering does not hold, which reduces the number of susceptible nodes adjacent to an infected large spreader $i$ because its neighbors are likely to have already been infected by $i$'s own infecting node. Both these effects lower the basic reproduction number $R_0$ below the theoretical value given by equation \eqref{EQ:R_0}.

\section{Conclusion}
Considering the problem of epidemic spreading of an infectious disease with a finite asymptomatic transmission rate, such as the current epidemics caused by the SARS-CoV-2, we have introduced a combined infection model of nodes taking susceptible, infected, or recovered states with a recursive contact tracing algorithm for quarantining, equivalent to an app used by a network's nodes to stop a pandemic in our model.\\
We have calculated the odds of an infected node being quarantined by the contact tracing algorithm, as well as the resulting theoretical critical values for the app usage rate 
above which an infection does not percolate through the network, and the minimum symptomatic rate beneath which a disease cannot be stopped, depending on the algorithm's recursion depth, the disease's basic reproduction number, and the contact network's underlying degree distribution.\\

We found that the critical app adoption rate and critical symptomatic rate are both significantly lower for an algorithm using recursive contact tracing, even with a low recursion depth, than for the classically employed, non-recursive method of direct contact tracing. In fact, any disease with a finite symptomatic rate and arbitrary basic reproduction number can be stopped if the app usage rate and recursion depth are large enough, meaning that recursive contact tracing can be an effective method for controlling diseases with large asymptomatic transmission rates which could not have been stopped with previous contact tracing methods.\\ 

Our critical app adoption rate of over 95\% may seem unusually high at first glance compared to some other results \cite{eames2003, hellewell2020, ferretti2020, hernandez2020}, with other estimates generally lying between 56\% and 95\% \cite{braithwaite2020}; however, this is simply caused by our model's harsh assumptions, such as a very high basic reproduction number $R_0=3$, a relatively high asymptomatic rate of 50\%, no infection prevention measures, such as random testing or social distancing, apart from contact tracing, no distinguishing between the infectivity of symptomatic and asymptomatic disease carriers (symptomatic carriers are often assumed to self-quarantine and therefore infect fewer people), and a lack of manual contact tracing even for symptomatic infected individuals who are not using the app. Our results are comparable to those of other models making harsh assumptions \cite{bulchandani2020, xia2020, lambert2020}.\\

Further, we found that, while higher recursion depths can stop diseases with a high asymptomatic rate, for low asymptomatic rates, recursion depths higher than one show very little improvement in the critical app usage rate while falsely quarantining more uninfected nodes, implying that for such diseases recursion depths larger than one are mostly not useful.\\
Also, the contact network's degree distribution was shown to have little impact on these critical values, so recursive contact tracing is not only viable for Erd\H{o}s-R\'enyi graphs, as tested in previous studies, but also for more realistic scale-free-like networks, i.e. scale-free networks with a cutoff.\\
We have ensured the accuracy of our theoretical calculations using simulations on infection trees and networks with different degree distributions. We found very good agreement between our calculations and simulations for any degree distribution on infection trees and for Erd\H{o}s-R\'enyi networks. For Barab\'asi-Albert networks, the simulation's critical values lay below the calculated ones because quarantining the most connected nodes quickly changes the network's degree distribution and because the effect of clustering, as highly connected nodes in Barab\'asi-Albert networks are likely to be connected to each other, was not considered in the calculations.\\

The calculations presented here are viable for a simple model, but we believe that the qualitative conclusions should be applicable to the real world as well. Future research should expand this simple model to be more realistic and possibly fit the infection profiles of real diseases, as well as consider the effect of clustering on the model's critical values. 

\bibliography{references}

\begin{thebibliography}{46}%
\makeatletter
\providecommand \@ifxundefined [1]{%
 \@ifx{#1\undefined}
}%
\providecommand \@ifnum [1]{%
 \ifnum #1\expandafter \@firstoftwo
 \else \expandafter \@secondoftwo
 \fi
}%
\providecommand \@ifx [1]{%
 \ifx #1\expandafter \@firstoftwo
 \else \expandafter \@secondoftwo
 \fi
}%
\providecommand \natexlab [1]{#1}%
\providecommand \enquote  [1]{``#1''}%
\providecommand \bibnamefont  [1]{#1}%
\providecommand \bibfnamefont [1]{#1}%
\providecommand \citenamefont [1]{#1}%
\providecommand \href@noop [0]{\@secondoftwo}%
\providecommand \href [0]{\begingroup \@sanitize@url \@href}%
\providecommand \@href[1]{\@@startlink{#1}\@@href}%
\providecommand \@@href[1]{\endgroup#1\@@endlink}%
\providecommand \@sanitize@url [0]{\catcode `\\12\catcode `\$12\catcode
  `\&12\catcode `\#12\catcode `\^12\catcode `\_12\catcode `\%12\relax}%
\providecommand \@@startlink[1]{}%
\providecommand \@@endlink[0]{}%
\providecommand \url  [0]{\begingroup\@sanitize@url \@url }%
\providecommand \@url [1]{\endgroup\@href {#1}{\urlprefix }}%
\providecommand \urlprefix  [0]{URL }%
\providecommand \Eprint [0]{\href }%
\providecommand \doibase [0]{http://dx.doi.org/}%
\providecommand \selectlanguage [0]{\@gobble}%
\providecommand \bibinfo  [0]{\@secondoftwo}%
\providecommand \bibfield  [0]{\@secondoftwo}%
\providecommand \translation [1]{[#1]}%
\providecommand \BibitemOpen [0]{}%
\providecommand \bibitemStop [0]{}%
\providecommand \bibitemNoStop [0]{.\EOS\space}%
\providecommand \EOS [0]{\spacefactor3000\relax}%
\providecommand \BibitemShut  [1]{\csname bibitem#1\endcsname}%
\let\auto@bib@innerbib\@empty
\bibitem [{\citenamefont {Franchini}\ \emph {et~al.}(2020)\citenamefont
  {Franchini}, \citenamefont {Auxilia}, \citenamefont {Galimberti},
  \citenamefont {Piga}, \citenamefont {Castaldi},\ and\ \citenamefont
  {Porro}}]{franchini2020}%
  \BibitemOpen
  \bibfield  {author} {\bibinfo {author} {\bibfnamefont {A.~F.}\ \bibnamefont
  {Franchini}}, \bibinfo {author} {\bibfnamefont {F.}~\bibnamefont {Auxilia}},
  \bibinfo {author} {\bibfnamefont {P.~M.}\ \bibnamefont {Galimberti}},
  \bibinfo {author} {\bibfnamefont {M.~A.}\ \bibnamefont {Piga}}, \bibinfo
  {author} {\bibfnamefont {S.}~\bibnamefont {Castaldi}}, \ and\ \bibinfo
  {author} {\bibfnamefont {A.}~\bibnamefont {Porro}},\ }\href@noop {}
  {\bibfield  {journal} {\bibinfo  {journal} {Acta bio-medica: Atenei
  Parmensis}\ }\textbf {\bibinfo {volume} {91}},\ \bibinfo {pages} {245}
  (\bibinfo {year} {2020})}\BibitemShut {NoStop}%
\bibitem [{\citenamefont {Wheelock}\ \emph {et~al.}(2020)\citenamefont
  {Wheelock} \emph {et~al.}}]{wheelock2020}%
  \BibitemOpen
  \bibfield  {author} {\bibinfo {author} {\bibfnamefont {D.~C.}\ \bibnamefont
  {Wheelock}} \emph {et~al.},\ }\href@noop {} {\bibfield  {journal} {\bibinfo
  {journal} {Economic Synopses}\ } (\bibinfo {year} {2020})}\BibitemShut
  {NoStop}%
\bibitem [{\citenamefont {Kretzschmar}\ \emph {et~al.}(1996)\citenamefont
  {Kretzschmar}, \citenamefont {van Duynhoven},\ and\ \citenamefont
  {Severijnen}}]{kretzschmar1996}%
  \BibitemOpen
  \bibfield  {author} {\bibinfo {author} {\bibfnamefont {M.}~\bibnamefont
  {Kretzschmar}}, \bibinfo {author} {\bibfnamefont {Y.~T.}\ \bibnamefont {van
  Duynhoven}}, \ and\ \bibinfo {author} {\bibfnamefont {A.~J.}\ \bibnamefont
  {Severijnen}},\ }\href@noop {} {\bibfield  {journal} {\bibinfo  {journal}
  {American Journal of Epidemiology}\ }\textbf {\bibinfo {volume} {144}},\
  \bibinfo {pages} {306} (\bibinfo {year} {1996})}\BibitemShut {NoStop}%
\bibitem [{\citenamefont {Eichner}(2003)}]{eichner2003}%
  \BibitemOpen
  \bibfield  {author} {\bibinfo {author} {\bibfnamefont {M.}~\bibnamefont
  {Eichner}},\ }\href@noop {} {\bibfield  {journal} {\bibinfo  {journal}
  {American journal of epidemiology}\ }\textbf {\bibinfo {volume} {158}},\
  \bibinfo {pages} {118} (\bibinfo {year} {2003})}\BibitemShut {NoStop}%
\bibitem [{\citenamefont {Eames}\ and\ \citenamefont
  {Keeling}(2003)}]{eames2003}%
  \BibitemOpen
  \bibfield  {author} {\bibinfo {author} {\bibfnamefont {K.~T.}\ \bibnamefont
  {Eames}}\ and\ \bibinfo {author} {\bibfnamefont {M.~J.}\ \bibnamefont
  {Keeling}},\ }\href@noop {} {\bibfield  {journal} {\bibinfo  {journal}
  {Proceedings of the Royal Society of London. Series B: Biological Sciences}\
  }\textbf {\bibinfo {volume} {270}},\ \bibinfo {pages} {2565} (\bibinfo {year}
  {2003})}\BibitemShut {NoStop}%
\bibitem [{\citenamefont {Fraser}\ \emph {et~al.}(2004)\citenamefont {Fraser},
  \citenamefont {Riley}, \citenamefont {Anderson},\ and\ \citenamefont
  {Ferguson}}]{fraser2004}%
  \BibitemOpen
  \bibfield  {author} {\bibinfo {author} {\bibfnamefont {C.}~\bibnamefont
  {Fraser}}, \bibinfo {author} {\bibfnamefont {S.}~\bibnamefont {Riley}},
  \bibinfo {author} {\bibfnamefont {R.~M.}\ \bibnamefont {Anderson}}, \ and\
  \bibinfo {author} {\bibfnamefont {N.~M.}\ \bibnamefont {Ferguson}},\
  }\href@noop {} {\bibfield  {journal} {\bibinfo  {journal} {Proceedings of the
  National Academy of Sciences}\ }\textbf {\bibinfo {volume} {101}},\ \bibinfo
  {pages} {6146} (\bibinfo {year} {2004})}\BibitemShut {NoStop}%
\bibitem [{\citenamefont {Kiss}\ \emph {et~al.}(2005)\citenamefont {Kiss},
  \citenamefont {Green},\ and\ \citenamefont {Kao}}]{kiss2005}%
  \BibitemOpen
  \bibfield  {author} {\bibinfo {author} {\bibfnamefont {I.~Z.}\ \bibnamefont
  {Kiss}}, \bibinfo {author} {\bibfnamefont {D.~M.}\ \bibnamefont {Green}}, \
  and\ \bibinfo {author} {\bibfnamefont {R.~R.}\ \bibnamefont {Kao}},\
  }\href@noop {} {\bibfield  {journal} {\bibinfo  {journal} {Proceedings of the
  Royal Society B: Biological Sciences}\ }\textbf {\bibinfo {volume} {272}},\
  \bibinfo {pages} {1407} (\bibinfo {year} {2005})}\BibitemShut {NoStop}%
\bibitem [{\citenamefont {M{\"u}ller}\ \emph {et~al.}(2000)\citenamefont
  {M{\"u}ller}, \citenamefont {Kretzschmar},\ and\ \citenamefont
  {Dietz}}]{muller2000}%
  \BibitemOpen
  \bibfield  {author} {\bibinfo {author} {\bibfnamefont {J.}~\bibnamefont
  {M{\"u}ller}}, \bibinfo {author} {\bibfnamefont {M.}~\bibnamefont
  {Kretzschmar}}, \ and\ \bibinfo {author} {\bibfnamefont {K.}~\bibnamefont
  {Dietz}},\ }\href@noop {} {\bibfield  {journal} {\bibinfo  {journal}
  {Mathematical biosciences}\ }\textbf {\bibinfo {volume} {164}},\ \bibinfo
  {pages} {39} (\bibinfo {year} {2000})}\BibitemShut {NoStop}%
\bibitem [{\citenamefont {Klinkenberg}\ \emph {et~al.}(2006)\citenamefont
  {Klinkenberg}, \citenamefont {Fraser},\ and\ \citenamefont
  {Heesterbeek}}]{klinkenberg2006}%
  \BibitemOpen
  \bibfield  {author} {\bibinfo {author} {\bibfnamefont {D.}~\bibnamefont
  {Klinkenberg}}, \bibinfo {author} {\bibfnamefont {C.}~\bibnamefont {Fraser}},
  \ and\ \bibinfo {author} {\bibfnamefont {H.}~\bibnamefont {Heesterbeek}},\
  }\href@noop {} {\bibfield  {journal} {\bibinfo  {journal} {PloS one}\
  }\textbf {\bibinfo {volume} {1}},\ \bibinfo {pages} {e12} (\bibinfo {year}
  {2006})}\BibitemShut {NoStop}%
\bibitem [{\citenamefont {Yu}\ and\ \citenamefont {Yang}(2020)}]{yu2020}%
  \BibitemOpen
  \bibfield  {author} {\bibinfo {author} {\bibfnamefont {X.}~\bibnamefont
  {Yu}}\ and\ \bibinfo {author} {\bibfnamefont {R.}~\bibnamefont {Yang}},\
  }\href@noop {} {\bibfield  {journal} {\bibinfo  {journal} {Influenza and
  Other Respiratory Viruses}\ } (\bibinfo {year} {2020})}\BibitemShut {NoStop}%
\bibitem [{\citenamefont {Pribylova}\ and\ \citenamefont
  {Hajnova}(2020)}]{pribylova2020}%
  \BibitemOpen
  \bibfield  {author} {\bibinfo {author} {\bibfnamefont {L.}~\bibnamefont
  {Pribylova}}\ and\ \bibinfo {author} {\bibfnamefont {V.}~\bibnamefont
  {Hajnova}},\ }\href@noop {} {\bibfield  {journal} {\bibinfo  {journal} {arXiv
  preprint arXiv:2004.02601}\ } (\bibinfo {year} {2020})}\BibitemShut {NoStop}%
\bibitem [{\citenamefont {Khailaie}\ \emph {et~al.}(2020)\citenamefont
  {Khailaie}, \citenamefont {Mitra}, \citenamefont {Bandyopadhyay},
  \citenamefont {Schips}, \citenamefont {Mascheroni}, \citenamefont {Vanella},
  \citenamefont {Lange}, \citenamefont {Binder},\ and\ \citenamefont
  {Meyer-Hermann}}]{khailaie2020}%
  \BibitemOpen
  \bibfield  {author} {\bibinfo {author} {\bibfnamefont {S.}~\bibnamefont
  {Khailaie}}, \bibinfo {author} {\bibfnamefont {T.}~\bibnamefont {Mitra}},
  \bibinfo {author} {\bibfnamefont {A.}~\bibnamefont {Bandyopadhyay}}, \bibinfo
  {author} {\bibfnamefont {M.}~\bibnamefont {Schips}}, \bibinfo {author}
  {\bibfnamefont {P.}~\bibnamefont {Mascheroni}}, \bibinfo {author}
  {\bibfnamefont {P.}~\bibnamefont {Vanella}}, \bibinfo {author} {\bibfnamefont
  {B.}~\bibnamefont {Lange}}, \bibinfo {author} {\bibfnamefont
  {S.}~\bibnamefont {Binder}}, \ and\ \bibinfo {author} {\bibfnamefont
  {M.}~\bibnamefont {Meyer-Hermann}},\ }\href@noop {} {\bibfield  {journal}
  {\bibinfo  {journal} {medRxiv}\ } (\bibinfo {year} {2020})}\BibitemShut
  {NoStop}%
\bibitem [{\citenamefont {Okolie}\ and\ \citenamefont
  {M{\"u}ller}(2020)}]{okolie2020}%
  \BibitemOpen
  \bibfield  {author} {\bibinfo {author} {\bibfnamefont {A.}~\bibnamefont
  {Okolie}}\ and\ \bibinfo {author} {\bibfnamefont {J.}~\bibnamefont
  {M{\"u}ller}},\ }\href@noop {} {\bibfield  {journal} {\bibinfo  {journal}
  {Mathematical Biosciences}\ }\textbf {\bibinfo {volume} {321}},\ \bibinfo
  {pages} {108320} (\bibinfo {year} {2020})}\BibitemShut {NoStop}%
\bibitem [{\citenamefont {Bulchandani}\ \emph {et~al.}(2020)\citenamefont
  {Bulchandani}, \citenamefont {Shivam}, \citenamefont {Moudgalya},\ and\
  \citenamefont {Sondhi}}]{bulchandani2020}%
  \BibitemOpen
  \bibfield  {author} {\bibinfo {author} {\bibfnamefont {V.~B.}\ \bibnamefont
  {Bulchandani}}, \bibinfo {author} {\bibfnamefont {S.}~\bibnamefont {Shivam}},
  \bibinfo {author} {\bibfnamefont {S.}~\bibnamefont {Moudgalya}}, \ and\
  \bibinfo {author} {\bibfnamefont {S.}~\bibnamefont {Sondhi}},\ }\href@noop {}
  {\bibfield  {journal} {\bibinfo  {journal} {arXiv preprint arXiv:2004.07237}\
  } (\bibinfo {year} {2020})}\BibitemShut {NoStop}%
\bibitem [{\citenamefont {Kojaku}\ \emph {et~al.}(2020)\citenamefont {Kojaku},
  \citenamefont {H{\'e}bert-Dufresne},\ and\ \citenamefont {Ahn}}]{kojaku2020}%
  \BibitemOpen
  \bibfield  {author} {\bibinfo {author} {\bibfnamefont {S.}~\bibnamefont
  {Kojaku}}, \bibinfo {author} {\bibfnamefont {L.}~\bibnamefont
  {H{\'e}bert-Dufresne}}, \ and\ \bibinfo {author} {\bibfnamefont {Y.-Y.}\
  \bibnamefont {Ahn}},\ }\href@noop {} {\bibfield  {journal} {\bibinfo
  {journal} {arXiv preprint arXiv:2005.02362}\ } (\bibinfo {year}
  {2020})}\BibitemShut {NoStop}%
\bibitem [{\citenamefont {Lambert}(2020)}]{lambert2020}%
  \BibitemOpen
  \bibfield  {author} {\bibinfo {author} {\bibfnamefont {A.}~\bibnamefont
  {Lambert}},\ }\href@noop {} {\bibfield  {journal} {\bibinfo  {journal}
  {medRxiv}\ } (\bibinfo {year} {2020})}\BibitemShut {NoStop}%
\bibitem [{\citenamefont {Barlow}(2020)}]{barlow2020}%
  \BibitemOpen
  \bibfield  {author} {\bibinfo {author} {\bibfnamefont {M.}~\bibnamefont
  {Barlow}},\ }\href@noop {} {\bibfield  {journal} {\bibinfo  {journal} {arXiv
  preprint arXiv:2007.16182}\ } (\bibinfo {year} {2020})}\BibitemShut {NoStop}%
\bibitem [{\citenamefont {Endo}\ \emph {et~al.}(2020)\citenamefont {Endo},
  \citenamefont {Leclerc}, \citenamefont {Knight}, \citenamefont {Medley},
  \citenamefont {Atkins}, \citenamefont {Funk}, \citenamefont {Kucharski} \emph
  {et~al.}}]{endo2020}%
  \BibitemOpen
  \bibfield  {author} {\bibinfo {author} {\bibfnamefont {A.}~\bibnamefont
  {Endo}}, \bibinfo {author} {\bibfnamefont {Q.~J.}\ \bibnamefont {Leclerc}},
  \bibinfo {author} {\bibfnamefont {G.~M.}\ \bibnamefont {Knight}}, \bibinfo
  {author} {\bibfnamefont {G.~F.}\ \bibnamefont {Medley}}, \bibinfo {author}
  {\bibfnamefont {K.~E.}\ \bibnamefont {Atkins}}, \bibinfo {author}
  {\bibfnamefont {S.}~\bibnamefont {Funk}}, \bibinfo {author} {\bibfnamefont
  {A.~J.}\ \bibnamefont {Kucharski}},  \emph {et~al.},\ }\href@noop {}
  {\bibfield  {journal} {\bibinfo  {journal} {medRxiv}\ } (\bibinfo {year}
  {2020})}\BibitemShut {NoStop}%
\bibitem [{\citenamefont {Faggian}\ \emph {et~al.}(2020)\citenamefont
  {Faggian}, \citenamefont {Urbani},\ and\ \citenamefont
  {Zanotto}}]{faggian2020}%
  \BibitemOpen
  \bibfield  {author} {\bibinfo {author} {\bibfnamefont {M.}~\bibnamefont
  {Faggian}}, \bibinfo {author} {\bibfnamefont {M.}~\bibnamefont {Urbani}}, \
  and\ \bibinfo {author} {\bibfnamefont {L.}~\bibnamefont {Zanotto}},\
  }\href@noop {} {\bibfield  {journal} {\bibinfo  {journal} {arXiv preprint
  arXiv:2003.10222}\ } (\bibinfo {year} {2020})}\BibitemShut {NoStop}%
\bibitem [{\citenamefont {Hellewell}\ \emph {et~al.}(2020)\citenamefont
  {Hellewell}, \citenamefont {Abbott}, \citenamefont {Gimma}, \citenamefont
  {Bosse}, \citenamefont {Jarvis}, \citenamefont {Russell}, \citenamefont
  {Munday}, \citenamefont {Kucharski}, \citenamefont {Edmunds}, \citenamefont
  {Sun} \emph {et~al.}}]{hellewell2020}%
  \BibitemOpen
  \bibfield  {author} {\bibinfo {author} {\bibfnamefont {J.}~\bibnamefont
  {Hellewell}}, \bibinfo {author} {\bibfnamefont {S.}~\bibnamefont {Abbott}},
  \bibinfo {author} {\bibfnamefont {A.}~\bibnamefont {Gimma}}, \bibinfo
  {author} {\bibfnamefont {N.~I.}\ \bibnamefont {Bosse}}, \bibinfo {author}
  {\bibfnamefont {C.~I.}\ \bibnamefont {Jarvis}}, \bibinfo {author}
  {\bibfnamefont {T.~W.}\ \bibnamefont {Russell}}, \bibinfo {author}
  {\bibfnamefont {J.~D.}\ \bibnamefont {Munday}}, \bibinfo {author}
  {\bibfnamefont {A.~J.}\ \bibnamefont {Kucharski}}, \bibinfo {author}
  {\bibfnamefont {W.~J.}\ \bibnamefont {Edmunds}}, \bibinfo {author}
  {\bibfnamefont {F.}~\bibnamefont {Sun}},  \emph {et~al.},\ }\href@noop {}
  {\bibfield  {journal} {\bibinfo  {journal} {The Lancet Global Health}\ }
  (\bibinfo {year} {2020})}\BibitemShut {NoStop}%
\bibitem [{\citenamefont {Hinch}\ \emph {et~al.}(2020)\citenamefont {Hinch},
  \citenamefont {Probert}, \citenamefont {Nurtay}, \citenamefont {Kendall},
  \citenamefont {Wymant}, \citenamefont {Hall},\ and\ \citenamefont
  {Fraser}}]{hinch2020}%
  \BibitemOpen
  \bibfield  {author} {\bibinfo {author} {\bibfnamefont {R.}~\bibnamefont
  {Hinch}}, \bibinfo {author} {\bibfnamefont {W.}~\bibnamefont {Probert}},
  \bibinfo {author} {\bibfnamefont {A.}~\bibnamefont {Nurtay}}, \bibinfo
  {author} {\bibfnamefont {M.}~\bibnamefont {Kendall}}, \bibinfo {author}
  {\bibfnamefont {C.}~\bibnamefont {Wymant}}, \bibinfo {author} {\bibfnamefont
  {M.}~\bibnamefont {Hall}}, \ and\ \bibinfo {author} {\bibfnamefont
  {C.}~\bibnamefont {Fraser}},\ }\href@noop {} {\bibfield  {journal} {\bibinfo
  {journal} {en. In:(Apr. 2020). Available here. url: https://github.
  com/BDI-pathogens/covid-19\_instant\_tracing/blob/master/Report}\ } (\bibinfo
  {year} {2020})}\BibitemShut {NoStop}%
\bibitem [{\citenamefont {Kim}\ and\ \citenamefont {Paul}(2020)}]{kim2020}%
  \BibitemOpen
  \bibfield  {author} {\bibinfo {author} {\bibfnamefont {H.}~\bibnamefont
  {Kim}}\ and\ \bibinfo {author} {\bibfnamefont {A.}~\bibnamefont {Paul}},\
  }\href@noop {} {\bibfield  {journal} {\bibinfo  {journal} {arXiv preprint
  arXiv:2004.10762}\ } (\bibinfo {year} {2020})}\BibitemShut {NoStop}%
\bibitem [{\citenamefont {Xia}\ and\ \citenamefont {Lee}(2020)}]{xia2020}%
  \BibitemOpen
  \bibfield  {author} {\bibinfo {author} {\bibfnamefont {Y.}~\bibnamefont
  {Xia}}\ and\ \bibinfo {author} {\bibfnamefont {G.}~\bibnamefont {Lee}},\
  }\href@noop {} {\bibfield  {journal} {\bibinfo  {journal} {arXiv preprint
  arXiv:2004.12576}\ } (\bibinfo {year} {2020})}\BibitemShut {NoStop}%
\bibitem [{\citenamefont {Ferretti}\ \emph {et~al.}(2020)\citenamefont
  {Ferretti}, \citenamefont {Wymant}, \citenamefont {Kendall}, \citenamefont
  {Zhao}, \citenamefont {Nurtay}, \citenamefont {Abeler-D{\"o}rner},
  \citenamefont {Parker}, \citenamefont {Bonsall},\ and\ \citenamefont
  {Fraser}}]{ferretti2020}%
  \BibitemOpen
  \bibfield  {author} {\bibinfo {author} {\bibfnamefont {L.}~\bibnamefont
  {Ferretti}}, \bibinfo {author} {\bibfnamefont {C.}~\bibnamefont {Wymant}},
  \bibinfo {author} {\bibfnamefont {M.}~\bibnamefont {Kendall}}, \bibinfo
  {author} {\bibfnamefont {L.}~\bibnamefont {Zhao}}, \bibinfo {author}
  {\bibfnamefont {A.}~\bibnamefont {Nurtay}}, \bibinfo {author} {\bibfnamefont
  {L.}~\bibnamefont {Abeler-D{\"o}rner}}, \bibinfo {author} {\bibfnamefont
  {M.}~\bibnamefont {Parker}}, \bibinfo {author} {\bibfnamefont
  {D.}~\bibnamefont {Bonsall}}, \ and\ \bibinfo {author} {\bibfnamefont
  {C.}~\bibnamefont {Fraser}},\ }\href@noop {} {\bibfield  {journal} {\bibinfo
  {journal} {Science}\ }\textbf {\bibinfo {volume} {368}} (\bibinfo {year}
  {2020})}\BibitemShut {NoStop}%
\bibitem [{\citenamefont {McLachlan}\ \emph {et~al.}(2020)\citenamefont
  {McLachlan}, \citenamefont {Lucas}, \citenamefont {Dube}, \citenamefont
  {McLachlan}, \citenamefont {Hitman}, \citenamefont {Osman},\ and\
  \citenamefont {Fenton}}]{mclachlan2020}%
  \BibitemOpen
  \bibfield  {author} {\bibinfo {author} {\bibfnamefont {S.}~\bibnamefont
  {McLachlan}}, \bibinfo {author} {\bibfnamefont {P.}~\bibnamefont {Lucas}},
  \bibinfo {author} {\bibfnamefont {K.}~\bibnamefont {Dube}}, \bibinfo {author}
  {\bibfnamefont {G.}~\bibnamefont {McLachlan}}, \bibinfo {author}
  {\bibfnamefont {G.}~\bibnamefont {Hitman}}, \bibinfo {author} {\bibfnamefont
  {M.}~\bibnamefont {Osman}}, \ and\ \bibinfo {author} {\bibfnamefont
  {N.}~\bibnamefont {Fenton}},\ }\href@noop {} {\enquote {\bibinfo {title} {The
  fundamental limitations of covid-19 contact tracing methods and how to
  resolve them with a bayesian network approach},}\ } (\bibinfo {year}
  {2020})\BibitemShut {NoStop}%
\bibitem [{\citenamefont {Hern{\'a}ndez-Orallo}\ \emph
  {et~al.}(2020)\citenamefont {Hern{\'a}ndez-Orallo}, \citenamefont {Manzoni},
  \citenamefont {Calafate},\ and\ \citenamefont {Cano}}]{hernandez2020}%
  \BibitemOpen
  \bibfield  {author} {\bibinfo {author} {\bibfnamefont {E.}~\bibnamefont
  {Hern{\'a}ndez-Orallo}}, \bibinfo {author} {\bibfnamefont {P.}~\bibnamefont
  {Manzoni}}, \bibinfo {author} {\bibfnamefont {C.~T.}\ \bibnamefont
  {Calafate}}, \ and\ \bibinfo {author} {\bibfnamefont {J.-C.}\ \bibnamefont
  {Cano}},\ }\href@noop {} {\bibfield  {journal} {\bibinfo  {journal} {IEEE
  Access}\ } (\bibinfo {year} {2020})}\BibitemShut {NoStop}%
\bibitem [{\citenamefont {Prasse}\ and\ \citenamefont
  {Van~Mieghem}(2020)}]{prasse2020}%
  \BibitemOpen
  \bibfield  {author} {\bibinfo {author} {\bibfnamefont {B.}~\bibnamefont
  {Prasse}}\ and\ \bibinfo {author} {\bibfnamefont {P.}~\bibnamefont
  {Van~Mieghem}},\ }\href@noop {} {\bibfield  {journal} {\bibinfo  {journal}
  {arXiv preprint arXiv:2006.14285}\ } (\bibinfo {year} {2020})}\BibitemShut
  {NoStop}%
\bibitem [{\citenamefont {Ho}\ \emph {et~al.}(2020)\citenamefont {Ho},
  \citenamefont {Chen}, \citenamefont {Hung}, \citenamefont {Huang},
  \citenamefont {Po}, \citenamefont {Chan}, \citenamefont {Yang}, \citenamefont
  {Tu}, \citenamefont {Liu},\ and\ \citenamefont {Fang}}]{ho2020}%
  \BibitemOpen
  \bibfield  {author} {\bibinfo {author} {\bibfnamefont {Y.-C.}\ \bibnamefont
  {Ho}}, \bibinfo {author} {\bibfnamefont {Y.-H.}\ \bibnamefont {Chen}},
  \bibinfo {author} {\bibfnamefont {S.-H.}\ \bibnamefont {Hung}}, \bibinfo
  {author} {\bibfnamefont {C.-H.}\ \bibnamefont {Huang}}, \bibinfo {author}
  {\bibfnamefont {P.}~\bibnamefont {Po}}, \bibinfo {author} {\bibfnamefont
  {C.-H.}\ \bibnamefont {Chan}}, \bibinfo {author} {\bibfnamefont {D.-K.}\
  \bibnamefont {Yang}}, \bibinfo {author} {\bibfnamefont {Y.-C.}\ \bibnamefont
  {Tu}}, \bibinfo {author} {\bibfnamefont {T.-L.}\ \bibnamefont {Liu}}, \ and\
  \bibinfo {author} {\bibfnamefont {C.-T.}\ \bibnamefont {Fang}},\ }\href@noop
  {} {\bibfield  {journal} {\bibinfo  {journal} {arXiv preprint
  arXiv:2006.16611}\ } (\bibinfo {year} {2020})}\BibitemShut {NoStop}%
\bibitem [{\citenamefont {Cencetti}\ \emph {et~al.}(2020)\citenamefont
  {Cencetti}, \citenamefont {Santin}, \citenamefont {Longa}, \citenamefont
  {Pigani}, \citenamefont {Barrat}, \citenamefont {Cattuto}, \citenamefont
  {Lehmann}, \citenamefont {Salathe},\ and\ \citenamefont
  {Lepri}}]{cencetti2020}%
  \BibitemOpen
  \bibfield  {author} {\bibinfo {author} {\bibfnamefont {G.}~\bibnamefont
  {Cencetti}}, \bibinfo {author} {\bibfnamefont {G.}~\bibnamefont {Santin}},
  \bibinfo {author} {\bibfnamefont {A.}~\bibnamefont {Longa}}, \bibinfo
  {author} {\bibfnamefont {E.}~\bibnamefont {Pigani}}, \bibinfo {author}
  {\bibfnamefont {A.}~\bibnamefont {Barrat}}, \bibinfo {author} {\bibfnamefont
  {C.}~\bibnamefont {Cattuto}}, \bibinfo {author} {\bibfnamefont
  {S.}~\bibnamefont {Lehmann}}, \bibinfo {author} {\bibfnamefont
  {M.}~\bibnamefont {Salathe}}, \ and\ \bibinfo {author} {\bibfnamefont
  {B.}~\bibnamefont {Lepri}},\ }\href@noop {} {\  (\bibinfo {year}
  {2020})}\BibitemShut {NoStop}%
\bibitem [{\citenamefont {Barrat}\ \emph {et~al.}(2020)\citenamefont {Barrat},
  \citenamefont {Cattuto}, \citenamefont {Kivel{\"a}}, \citenamefont
  {Lehmann},\ and\ \citenamefont {Saram{\"a}ki}}]{barrat2020}%
  \BibitemOpen
  \bibfield  {author} {\bibinfo {author} {\bibfnamefont {A.}~\bibnamefont
  {Barrat}}, \bibinfo {author} {\bibfnamefont {C.}~\bibnamefont {Cattuto}},
  \bibinfo {author} {\bibfnamefont {M.}~\bibnamefont {Kivel{\"a}}}, \bibinfo
  {author} {\bibfnamefont {S.}~\bibnamefont {Lehmann}}, \ and\ \bibinfo
  {author} {\bibfnamefont {J.}~\bibnamefont {Saram{\"a}ki}},\ }\href@noop {}
  {\bibfield  {journal} {\bibinfo  {journal} {medRxiv}\ } (\bibinfo {year}
  {2020})}\BibitemShut {NoStop}%
\bibitem [{\citenamefont {Grassberger}(1983)}]{grassberger1983}%
  \BibitemOpen
  \bibfield  {author} {\bibinfo {author} {\bibfnamefont {P.}~\bibnamefont
  {Grassberger}},\ }\href@noop {} {\bibfield  {journal} {\bibinfo  {journal}
  {Mathematical Biosciences}\ }\textbf {\bibinfo {volume} {63}},\ \bibinfo
  {pages} {157} (\bibinfo {year} {1983})}\BibitemShut {NoStop}%
\bibitem [{\citenamefont {Cardy}\ and\ \citenamefont
  {Grassberger}(1985)}]{cardy1985}%
  \BibitemOpen
  \bibfield  {author} {\bibinfo {author} {\bibfnamefont {J.~L.}\ \bibnamefont
  {Cardy}}\ and\ \bibinfo {author} {\bibfnamefont {P.}~\bibnamefont
  {Grassberger}},\ }\href@noop {} {\bibfield  {journal} {\bibinfo  {journal}
  {Journal of Physics A: Mathematical and General}\ }\textbf {\bibinfo {volume}
  {18}},\ \bibinfo {pages} {L267} (\bibinfo {year} {1985})}\BibitemShut
  {NoStop}%
\bibitem [{\citenamefont {Newman}\ and\ \citenamefont
  {Watts}(1999)}]{newman1999}%
  \BibitemOpen
  \bibfield  {author} {\bibinfo {author} {\bibfnamefont {M.~E.}\ \bibnamefont
  {Newman}}\ and\ \bibinfo {author} {\bibfnamefont {D.~J.}\ \bibnamefont
  {Watts}},\ }\href@noop {} {\bibfield  {journal} {\bibinfo  {journal}
  {Physical review E}\ }\textbf {\bibinfo {volume} {60}},\ \bibinfo {pages}
  {7332} (\bibinfo {year} {1999})}\BibitemShut {NoStop}%
\bibitem [{\citenamefont {Moore}\ and\ \citenamefont
  {Newman}(2000)}]{moore2000}%
  \BibitemOpen
  \bibfield  {author} {\bibinfo {author} {\bibfnamefont {C.}~\bibnamefont
  {Moore}}\ and\ \bibinfo {author} {\bibfnamefont {M.~E.}\ \bibnamefont
  {Newman}},\ }\href@noop {} {\bibfield  {journal} {\bibinfo  {journal}
  {Physical Review E}\ }\textbf {\bibinfo {volume} {61}},\ \bibinfo {pages}
  {5678} (\bibinfo {year} {2000})}\BibitemShut {NoStop}%
\bibitem [{\citenamefont {Pastor-Satorras}\ and\ \citenamefont
  {Vespignani}(2001{\natexlab{a}})}]{pastor2001}%
  \BibitemOpen
  \bibfield  {author} {\bibinfo {author} {\bibfnamefont {R.}~\bibnamefont
  {Pastor-Satorras}}\ and\ \bibinfo {author} {\bibfnamefont {A.}~\bibnamefont
  {Vespignani}},\ }\href@noop {} {\bibfield  {journal} {\bibinfo  {journal}
  {Physical Review E}\ }\textbf {\bibinfo {volume} {63}},\ \bibinfo {pages}
  {066117} (\bibinfo {year} {2001}{\natexlab{a}})}\BibitemShut {NoStop}%
\bibitem [{\citenamefont {Newman}(2002)}]{newman2002}%
  \BibitemOpen
  \bibfield  {author} {\bibinfo {author} {\bibfnamefont {M.~E.}\ \bibnamefont
  {Newman}},\ }\href@noop {} {\bibfield  {journal} {\bibinfo  {journal}
  {Physical review E}\ }\textbf {\bibinfo {volume} {66}},\ \bibinfo {pages}
  {016128} (\bibinfo {year} {2002})}\BibitemShut {NoStop}%
\bibitem [{\citenamefont {Warren}\ \emph {et~al.}(2002)\citenamefont {Warren},
  \citenamefont {Sander},\ and\ \citenamefont {Sokolov}}]{warren2002}%
  \BibitemOpen
  \bibfield  {author} {\bibinfo {author} {\bibfnamefont {C.~P.}\ \bibnamefont
  {Warren}}, \bibinfo {author} {\bibfnamefont {L.~M.}\ \bibnamefont {Sander}},
  \ and\ \bibinfo {author} {\bibfnamefont {I.~M.}\ \bibnamefont {Sokolov}},\
  }\href@noop {} {\bibfield  {journal} {\bibinfo  {journal} {Physical Review
  E}\ }\textbf {\bibinfo {volume} {66}},\ \bibinfo {pages} {056105} (\bibinfo
  {year} {2002})}\BibitemShut {NoStop}%
\bibitem [{\citenamefont {Keeling}\ and\ \citenamefont
  {Eames}(2005)}]{keeling2005}%
  \BibitemOpen
  \bibfield  {author} {\bibinfo {author} {\bibfnamefont {M.~J.}\ \bibnamefont
  {Keeling}}\ and\ \bibinfo {author} {\bibfnamefont {K.~T.}\ \bibnamefont
  {Eames}},\ }\href@noop {} {\bibfield  {journal} {\bibinfo  {journal} {Journal
  of the Royal Society Interface}\ }\textbf {\bibinfo {volume} {2}},\ \bibinfo
  {pages} {295} (\bibinfo {year} {2005})}\BibitemShut {NoStop}%
\bibitem [{\citenamefont {Meyers}(2007)}]{meyers2007}%
  \BibitemOpen
  \bibfield  {author} {\bibinfo {author} {\bibfnamefont {L.}~\bibnamefont
  {Meyers}},\ }\href@noop {} {\bibfield  {journal} {\bibinfo  {journal}
  {Bulletin of the American Mathematical Society}\ }\textbf {\bibinfo {volume}
  {44}},\ \bibinfo {pages} {63} (\bibinfo {year} {2007})}\BibitemShut {NoStop}%
\bibitem [{\citenamefont {Pastor-Satorras}\ \emph {et~al.}(2015)\citenamefont
  {Pastor-Satorras}, \citenamefont {Castellano}, \citenamefont {Van~Mieghem},\
  and\ \citenamefont {Vespignani}}]{pastor2015}%
  \BibitemOpen
  \bibfield  {author} {\bibinfo {author} {\bibfnamefont {R.}~\bibnamefont
  {Pastor-Satorras}}, \bibinfo {author} {\bibfnamefont {C.}~\bibnamefont
  {Castellano}}, \bibinfo {author} {\bibfnamefont {P.}~\bibnamefont
  {Van~Mieghem}}, \ and\ \bibinfo {author} {\bibfnamefont {A.}~\bibnamefont
  {Vespignani}},\ }\href@noop {} {\bibfield  {journal} {\bibinfo  {journal}
  {Reviews of modern physics}\ }\textbf {\bibinfo {volume} {87}},\ \bibinfo
  {pages} {925} (\bibinfo {year} {2015})}\BibitemShut {NoStop}%
\bibitem [{\citenamefont {Pastor-Satorras}\ and\ \citenamefont
  {Vespignani}(2001{\natexlab{b}})}]{pastor2001_2}%
  \BibitemOpen
  \bibfield  {author} {\bibinfo {author} {\bibfnamefont {R.}~\bibnamefont
  {Pastor-Satorras}}\ and\ \bibinfo {author} {\bibfnamefont {A.}~\bibnamefont
  {Vespignani}},\ }\href@noop {} {\bibfield  {journal} {\bibinfo  {journal}
  {Physical review letters}\ }\textbf {\bibinfo {volume} {86}},\ \bibinfo
  {pages} {3200} (\bibinfo {year} {2001}{\natexlab{b}})}\BibitemShut {NoStop}%
\bibitem [{\citenamefont {Madar}\ \emph {et~al.}(2004)\citenamefont {Madar},
  \citenamefont {Kalisky}, \citenamefont {Cohen}, \citenamefont {Ben-avraham},\
  and\ \citenamefont {Havlin}}]{madar2004}%
  \BibitemOpen
  \bibfield  {author} {\bibinfo {author} {\bibfnamefont {N.}~\bibnamefont
  {Madar}}, \bibinfo {author} {\bibfnamefont {T.}~\bibnamefont {Kalisky}},
  \bibinfo {author} {\bibfnamefont {R.}~\bibnamefont {Cohen}}, \bibinfo
  {author} {\bibfnamefont {D.}~\bibnamefont {Ben-avraham}}, \ and\ \bibinfo
  {author} {\bibfnamefont {S.}~\bibnamefont {Havlin}},\ }\href@noop {}
  {\bibfield  {journal} {\bibinfo  {journal} {The European Physical Journal B}\
  }\textbf {\bibinfo {volume} {38}},\ \bibinfo {pages} {269} (\bibinfo {year}
  {2004})}\BibitemShut {NoStop}%
\bibitem [{\citenamefont {Lloyd-Smith}\ \emph {et~al.}(2005)\citenamefont
  {Lloyd-Smith}, \citenamefont {Schreiber}, \citenamefont {Kopp},\ and\
  \citenamefont {Getz}}]{lloyd2005}%
  \BibitemOpen
  \bibfield  {author} {\bibinfo {author} {\bibfnamefont {J.~O.}\ \bibnamefont
  {Lloyd-Smith}}, \bibinfo {author} {\bibfnamefont {S.~J.}\ \bibnamefont
  {Schreiber}}, \bibinfo {author} {\bibfnamefont {P.~E.}\ \bibnamefont {Kopp}},
  \ and\ \bibinfo {author} {\bibfnamefont {W.~M.}\ \bibnamefont {Getz}},\
  }\href@noop {} {\bibfield  {journal} {\bibinfo  {journal} {Nature}\ }\textbf
  {\bibinfo {volume} {438}},\ \bibinfo {pages} {355} (\bibinfo {year}
  {2005})}\BibitemShut {NoStop}%
\bibitem [{\citenamefont {Castellano}\ and\ \citenamefont
  {Pastor-Satorras}(2010)}]{castellano2010}%
  \BibitemOpen
  \bibfield  {author} {\bibinfo {author} {\bibfnamefont {C.}~\bibnamefont
  {Castellano}}\ and\ \bibinfo {author} {\bibfnamefont {R.}~\bibnamefont
  {Pastor-Satorras}},\ }\href@noop {} {\bibfield  {journal} {\bibinfo
  {journal} {Physical review letters}\ }\textbf {\bibinfo {volume} {105}},\
  \bibinfo {pages} {218701} (\bibinfo {year} {2010})}\BibitemShut {NoStop}%
\bibitem [{\citenamefont {Meyers}\ \emph {et~al.}(2005)\citenamefont {Meyers},
  \citenamefont {Pourbohloul}, \citenamefont {Newman}, \citenamefont
  {Skowronski},\ and\ \citenamefont {Brunham}}]{meyers2005}%
  \BibitemOpen
  \bibfield  {author} {\bibinfo {author} {\bibfnamefont {L.~A.}\ \bibnamefont
  {Meyers}}, \bibinfo {author} {\bibfnamefont {B.}~\bibnamefont {Pourbohloul}},
  \bibinfo {author} {\bibfnamefont {M.~E.}\ \bibnamefont {Newman}}, \bibinfo
  {author} {\bibfnamefont {D.~M.}\ \bibnamefont {Skowronski}}, \ and\ \bibinfo
  {author} {\bibfnamefont {R.~C.}\ \bibnamefont {Brunham}},\ }\href@noop {}
  {\bibfield  {journal} {\bibinfo  {journal} {Journal of theoretical biology}\
  }\textbf {\bibinfo {volume} {232}},\ \bibinfo {pages} {71} (\bibinfo {year}
  {2005})}\BibitemShut {NoStop}%
\bibitem [{\citenamefont {Braithwaite}\ \emph {et~al.}(2020)\citenamefont
  {Braithwaite}, \citenamefont {Callender}, \citenamefont {Bullock},\ and\
  \citenamefont {Aldridge}}]{braithwaite2020}%
  \BibitemOpen
  \bibfield  {author} {\bibinfo {author} {\bibfnamefont {I.}~\bibnamefont
  {Braithwaite}}, \bibinfo {author} {\bibfnamefont {T.}~\bibnamefont
  {Callender}}, \bibinfo {author} {\bibfnamefont {M.}~\bibnamefont {Bullock}},
  \ and\ \bibinfo {author} {\bibfnamefont {R.~W.}\ \bibnamefont {Aldridge}},\
  }\href@noop {} {\bibfield  {journal} {\bibinfo  {journal} {medRxiv}\ }
  (\bibinfo {year} {2020})}\BibitemShut {NoStop}%
\end{thebibliography}%
\end{document}